\documentstyle[preprint,aps,epsf]{revtex}

\begin{document}
\vskip 2true cm
\title{
Molecular dynamics study of solvation effects on acid dissociation
in aprotic media }
\author{Daniel Laria\\ 
Departamento Qu\'{\i}mica de Reactores,\\ Comisi\'{o}n
Nacional de Energ\'{\i}a At\'{o}mica, 1429 Capital Federal, Argentina\\
Raymond Kapral\\
Chemical Physics Theory Group, Department of Chemistry,\\ University of
Toronto, Toronto, Ontario M5S 1A1, Canada\\
Dar\' \i o Estrin\\
INQUIMAE, Facultad de Ciencias Exactas y Naturales, Universidad
de Buenos Aires\\
Ciudad Universitaria, Pabell\'on II. 1428 Capital Federal,
Argentina\\
Giovanni Ciccotti\\
Dipartimento di Fisica, Universit\`{a} "La
Sapienza",\\ Piazzale Aldo Moro, 2, 00185 Roma, Italy\\}

\newpage

\maketitle
\begin{abstract}
Acid ionization in aprotic media is studied using Molecular Dynamics
techniques.  In particular,  models for HCl ionization in acetonitrile
and dimethylsulfoxide are investigated. The proton is treated quantum
mechanically using Feynman path integral methods and the remaining
molecules are treated classically. Quantum effects are shown to be essential 
for the proper treatment of the ionization. The potential of mean force 
is computed as a function
of the ion pair separation  and the local solvent structure
is examined. The computed dissociation constants in both solvents
differ by several orders of magnitude 
which are in reasonable agreement with experimental results. 
Solvent separated ion
pairs are found to exist in dimethylsulfoxide but not in acetonitrile.
Dissociation mechanisms in small clusters are also investigated.  
Solvent separated ion pairs persist even in aggregates composed of
rather few molecules, for instance, as few as thirty molecules. For
smaller clusters or for large ion pair separations cluster finite-size
effects come into play in a significant fashion.
\end{abstract}

\newpage

\section{Introduction}
The ionization of an acid, HA, in condensed phases
\begin{equation}
{\rm HA} \stackrel{{\rm S}}{\longrightarrow} {\rm H}^+({\rm S}) 
+{\rm A}^-({\rm S})\ 
\end{equation}
is clearly influenced by the nature of the solvent S
in which the ionization takes place. A molecular understanding of the processes
responsible for these solvent effects requires an analysis of the solvation
forces that bring about the ionization and examination of how 
these forces depend on the
properties of the solvent molecules. The analysis is complicated by the
fact that the proton is a quantum object and thus the theoretical
description must account for its quantum character in determining the
interactions between the proton and the solvent.

In this article, we study such acid ionization process in aprotic
media.  We have selected two simple prototypic molecular solvents
for this study: acetonitrile (ACN) and dimethylsulfoxide (DMSO).
While these two molecules have similar dipolar
properties they have markedly different effects on the ionization process; for
example, the dissociation constant for HCl in DMSO is several orders of
magnitude greater than that in  ACN.\cite{dis}
Furthermore, aprotic solvents have been chosen in
order to avoid complications due to cooperative proton motion in the
solvent (Grotthuss mechanism)\cite{grot} that 
may play a role when water is the solvent.\cite{hynes}

The main focus of this work is on the determination of the potential of
mean force as a function of the separation between the H$^+$ and Cl$^-$
ions. This allows us to characterize the equilibrium 
solvation structure; in particular, if solvent
separated ion pairs (SSIP)\cite{ssip} can exist in either solvent so that the
ionization process should be represented by
\begin{equation}
{\rm HCl} \longrightarrow {\rm H}^+ || {\rm Cl}^- \longrightarrow  
{\rm H^+(S) +Cl^-(S)} \ .
\end{equation}
Here H$^+ ||$ Cl$^-$ represents a metastable configuration corresponding
to a local minimum in the mean potential where solvent molecules reside
between the ions and the ions may not yet be considered to be 
completely free ions
in solution.\cite{ipsim}
Our calculations permit us to
determine properties of the solvated quantum proton and to study 
the differences in the 
solvation properties in both solvents.

Since the ionization process is an example of a
reaction that is strongly solvent influenced, it is also of interest to see how
the chemistry of ionization processes is changed in finite-size
systems. Consequently, we also examine molecular clusters with linear 
dimensions in the nanometer range to study how the ionization process 
changes as a function of the cluster size. 
It is well known that chemical reactions in clusters present some 
unique features due to the lack of translational symmetry;
for example, 
certain charge transfer reactions leading to neutral ion pairs take
place in confined systems only above a 
certain threshold cluster size.\cite{ct,ct1}
Acid dissociative processes occurring in aqueous clusters are
also known to exhibit similar behavior where cooperative
effects gradually  increase,  stabilizing the reaction
products as the number of the components in the cluster increases.\cite{cast} 
Consequently, it is interesting to extend  
bulk studies to cluster domains to unveil peculiarities of 
the mechanisms that govern dissociative processes
in aggregates containing small numbers of particles. 

The organization of the paper is as follows:
Section II gives details of the interaction potentials used in this work,
along with the molecular dynamics methods used to carry out the
simulations of the mean potential. Bulk phase and cluster results  are 
contained in Secs. III and IV, respectively.
The conclusions of this study are given in Sec. V.    

\section{System Definition}
\subsection{Solvent potential parameters}
The ACN [CH$_3$CN] and DMSO [(CH$_3$)$_2$SO] solvents considered in this study
were modeled as rigid molecules with the following characteristics:
ACN was comprised of three sites representing a united atom model for
the  CH$_3$ group, C and
N units in the linear molecule with separations $d$ between sites in
a molecule fixed by constraints\cite{shake} at the following values: 
$d$(C-CH$_3$)=1.46 \AA \   and $d$(N-CH$_3$)=2.63 \AA. 
The molecular dipole moment is
$\mu_{ACN}=4.14 D$. The molecules interact by site-site Lennard-Jones (LJ)
and Coulomb interactions. The LJ parameters and partial charges were
taken from Edwards et al.\cite{edwards} and  
interactions between identical sites are given by 
$\sigma_N=3.3$, $\sigma_C=3.4$ and $\sigma_{CH_3}=3.6$; $\epsilon_N=50$, 
$\epsilon_C=50$ and $\epsilon_{CH_3}=191$; $q_N=-0.398$, $q_C=0.129$ and 
$q_{CH_3}=0.269$. Here and below the $\sigma$ values are expressed in
\AA, the $\epsilon$ parameters in K and the partial charges in 
units of the electronic charge $e$. The usual arithmetic and geometrical 
means were used to determine respectively the $\sigma$ and $\epsilon$ 
parameters for the cross interactions.

DMSO was modeled as a rigid four-site molecule with
the following internuclear separations between constituents\cite{feder}:
$d$(S-CH$_3$)=1.799 \AA, $d$(S-O)=1.48 \AA, $d$(O-CH$_3$)=2.657 \AA  \   and
$d$(CH$_3$-CH$_3$)=2.685 \AA. The molecule is nonlinear with the following 
angles among the constituents: $\theta$(CH$_3$-S-CH$_3$)$=96^{\circ}34^{\prime}$, 
$\theta$(CH$_3$-S-O)$=106^{\circ}39^{\prime}$ 
and the angle that the OS bond makes with the CH$_3$-S-CH$_3$ plane is
$64^{\circ}30^{\prime}$. The molecular dipole moment is
$\mu_{DMSO}=4.42 D$.  The LJ parameters and partial charges were taken from
Rao and Singh\cite{rao} and for interactions between
the identical sites are given by 
$\sigma_O=3.3$, $\sigma_S=4.0$ and $\sigma_{CH_3}=4.03$; $\epsilon_O=33.2$, 
$\epsilon_S=101.7$ and $\epsilon_{CH_3}=80.5$; $q_O=-0.459$, $q_S=0.139$ and 
$q_{CH_3}=0.160$. Parameters for cross interactions were determined 
as described above.

\subsection{HCl and solvent-solute interactions}
We consider a simple model for the ionization of HCl in the above
solvents. In the gas phase the lowest energy potential energy
surface corresponds to dissociation into neutral atoms while in solution,
the molecule dissociates into ions due to solvation forces. A full quantum 
calculation of the potential energy surface appropriate 
for the ionization in solution must account for solvent degrees 
of freedom.  We considered a much simpler model. The interaction potential 
energy between the H and Cl atoms
leading to dissociation into ions  was determined
from a density functional calculation for a Cl$^-$ ion in the field of a
point positive charge a fixed distance away. 
By appropriate distribution of the basis functions describing the electronic
density of the complex in the neighborhood of the chlorine atom, the ``bare" 
molecule was artificially constrained to dissociate into Cl$^-$ and H$^+$ ions at 
large separations, while the charge density was allowed to redistribute at 
shorter separations. The computation was 
performed at the Generalized Gradient Approximation level\cite {dft1} 
 using a  flexible basis set for the Cl$^-$.\cite{dft2}  
Results of these calculations as a function of the interionic
distances $r$ were fitted   
using the following potential energy function
\begin{equation}
\label{vhcl}
V_{H^+Cl^-}(r)= A \  \exp (-\lambda r) -{e^2 \over r}\;,
\end{equation}
where $e$ is the electron charge,  
$A=1.273 \times 10^4$ kcal/mol and $\lambda=4.159$ \AA$^{-1}$. Note
that in this functional form the Coulomb interaction does not depend 
on the partial charges on the ions determined from the computed electronic
density. Nevertheless, in the interactions between the ions
and the solvent sites, the variation of the ionic charge with the distance
$r$ between the Cl$^-$ and H$^+$ ions was taken into account.
In this way, the changes in the
electronic density of the HCl molecule during  the dissociation
process was taken into account.
>From our density functional calculations, the
charge on the Cl$^-$ ion for $r \ge$ 1 \AA \  could be represented 
approximately by the formula
\begin{equation}
q_{CL}(r)=-1+0.905 \ e^{- \gamma r}\;,
\end{equation}
with $\gamma= 0.48$ \AA$^{-1}$ , $q_H(r)=-q_{Cl}(r)$. 
In performing the calculations, we have made no attempt to reproduce 
with (\ref{vhcl})  
the actual value of dissociation energy of HCl in $vacuo$ 
or the exact value of the equilibrium distance for the HCl 
chemical bond. In fact, our simple
approach based on density functional theory was used only as
a reasonable interpolative scheme between two well
defined electronic structures for reactants and products states: the covalent 
character of the
intramolecular bond at short interatomic distances and the ionic character 
of the dissociated species. Although this  is a crude approximation, 
it captures the importante features of the solvation effects in 
different environments, especially at the larger separations of interest 
in this study where the species are well approximated by H$^+$ and 
Cl$^-$ ions.

Using the charge density extracted from this
calculation, the interactions between the Cl$^-$ and H$^+$ ions and the
solvent molecules were taken to be site-site LJ plus Coulomb
interactions. The LJ parameters for these ions were taken from 
data on Cl$^-$ in H$_2$O given in Rossky et al.\cite{rossky} using 
the normal geometric and arithmetic means and the known parameters 
for SPC water. 
The values used in the present calculation are 
$\sigma_{Cl^-}=3.93$ and $\sigma_{H^+}=0.35$; $\epsilon_{Cl^-}=0.830$, 
and $\epsilon_{H^+}=0.155$ and,
as earlier, the LJ parameters for cross interactions between these ions
and the molecular solvent sites were determined from geometric and
arithmetic means.

\subsection{Simulation details}
The proton was treated quantum mechanically using Feynman's path
integral formulation of quantum mechanics.\cite{feyn} In this representation the
proton is represented by a ring polymer with harmonic bonds between the 
polymer monomers, which also interact with the other classical particles
in the system.\cite{feyn,chand} The effective potential of the system is 
\begin{equation}
V_{eff}=V_{cl}({\bf R}^N) +{P m_p \over 2 (\beta \hbar)^2} \sum_{i=1}^P 
({\bf r}_i -{\bf r}_{i+1})^2 +{1 \over P} V_p({\bf r}^P,{\bf R}^N)\;,
\end{equation}
where $\beta^{-1}$ is Boltzmann's constant times temperature, $m_p$ represents
the proton mass,
${\bf R}^N$ refers to the set of classical coordinates and 
${\bf r}^P$ to the proton monomer coordinates with $P$ the number of 
monomers $({\bf r}^{P+1}={\bf r}^1)$. 
Here $V_{cl}({\bf R}^N)$ is the potential energy of all
classical particles and $V_p({\bf r}^P,{\bf R}^N)$ is the interaction
energy between the proton monomers and the classical particles. 
We have taken $P=20$ in order to accurately represent the
proton but the results were checked against simulations with $P=40$ and
no significant changes in the equilibrium averages were seen.

Equilibrium averages were computed from time averages over a fictitious
molecular dynamics (MD) generated by the following Hamiltonian:
\begin{equation}
H=\sum_{i=1}^P {1 \over 2} m^* {\bf \dot{r}}_i^2 +\sum_{j=1}^N {1 \over 2} M_j
{\bf \dot{R}}_j^2 + V_{eff}\;,
\end{equation}
where $m^*$ is the fictitious mass assigned to the proton monomers. We
have taken $m^*=30$ atomic units. Given this choice the oscillation
period for the harmonic forces in the polymer is 
\begin{equation}
\tau_{osc}={2 \pi \over \omega}=
2 \pi \left({m^* \over  P m_p}\right)^{1/2} \beta 
\hbar\;.
\end{equation}
Using the values given above, at $T$=200 K and $T$=293 K
the period is $\tau_{osc}=207$ fs and $\tau_{osc}=109$ fs, respectively. 

The bulk phase molecular dynamics simulations were carried out on a system composed
of 142 solvent molecules plus the two ions confined to a periodic box
with sides $L$. The LJ interactions were cut off at half the size of
the simulation box. A quartic spline interpolation was used to make 
the the Coulomb interactions go smoothly to zero at $L/2$ 
in a 1 \AA \ region. The box sizes were $L$=23.2 \AA \ for ACN and 
$L$=25.78 \AA  \  
for DMSO giving densities of 52.2 cm$^3$/mol and 71.7 cm$^3$/mol, respectively. 
The bulk simulations were
carried out at constant temperature using Nos\'e dynamics.\cite{nose} 
Two Nos\'e themostats were used, one for the proton polymer and the other 
for the remaining classical particles. The temperatures of both thermostats were
fixed at 293 K. The MD integration was performed using the
Verlet algorithm\cite{verlet} with a time step of 2 fs. Note that with this 
time step there are
between 50 and 100 time steps per oscillation period of the proton
monomers.

The cluster results were obtained from time averages over constant
energy molecular dynamics simulations.
The clusters were
equilibrated for 50 ps using constant temperature (Nos\'e) 
molecular dynamics 
after this period the thermostat was switched off. 
Time averages were then determined from 4-5 ns constant
energy MD trajectories. Clusters with sizes ranging from $n=4$ to $n=30$ 
solvent molecules were studied. The average temperature was $200 \pm
20$ K for the small clusters with much smaller
fluctuations for the larger clusters.

\section{Acid Ionization  in Bulk Solvents}
Our study of  acid ionization will be primarily concerned  
with the change in character of the
ionization process, as reflected in the potential of mean force or free
energy for varying separation of the ions, as a function of the
different solvent environment. In particular, we will be interested
in describing $(i)$ 
the magnitude of the resulting free energy barriers for the dissociation
processes in both solvents and  $(ii)$ the main features that characterize 
the solvation structure of the complex as the reactive process takes place.  

\subsection{Potential of mean force}
The mean potential  as a function of the distance between the ions was
obtained from the mean force on the ion pair.\cite{path} The proton position was
defined as the centroid of the proton polymer,\cite{gillan,voth}
\begin{equation}
{\bf r}_c={1 \over P} \sum_{i=1}^P {\bf r}_i\;,
\end{equation}
and thus the distance between the Cl$^-$ and H$^+$ ions is 
\begin{equation}
r=|{\bf R}_{Cl}-{\bf r}_c|\;,
\end{equation}
with ${\bf R}_{Cl}$ the position of the Cl$^-$ ion.
The mean force at a fixed ion pair displacement $r$ was calculated as\cite{path}
\begin{equation}
F(r) =\langle \sum_{i=1}^{P} {\bf F}_{i} \cdot \hat{{\bf r}} \rangle_{r}
+ \frac{2}{\beta r}
\end{equation}
where ${\bf F}_{i}$ is the force exerted
by the Cl$^-$ and the solvent particles
upon the $i$-th proton monomer,
$\hat{\bf{r}}$ represents
the versor along the interionic distance and $\langle  .... \rangle_{r}$
represent time average computed using the constrained reaction 
coordinate dynamics ensemble.\cite{crcd} 
The mean potential $W(r)$ was then found by
integration of the mean force using
\begin{equation}
F(r)=-{d W(r) \over dr}\;.
\end{equation}

Figure~1 shows the potential of mean force for the ion pair in bulk 
ACN and DMSO solvents.
For reference the ``bare" HCl
potential in the absence of solvent is also shown. The potentials have
been shifted to correspond at 
their first minima. (We term this the contact ion pair state (CIP). In 
this case there is no distinction between the bound molecular species 
and the contact ion pair state.) 
We note the pronounced
differences between the ACN and DMSO solvation. For example, at
$r_{max}=4.0$ \AA, the maximum in the DMSO mean potential,
the curves differ by 
$\beta \Delta W(r_{max})=W_{ACN}(r_{max})-W_{DMSO}(r_{max}) =
10.9$.  From these curves,  we can easily extract 
the ratio of dissociation constants $K_d$  
\begin{equation}\label{rough}
{K_d(DMSO) \over K_d(ACN)} = \frac {  \left[
\int \ \Theta_{ACN}(r) \   e^{-\beta W_{ACN}(r)} \  {\rm d}{\bf r}\right]
 \ \ \left[
\int \ (1-\Theta_{DMSO}(r)) \   e^{-\beta W_{DMSO}(r)} \  {\rm d}{\bf r}
\right]^2
 }    { \left[
\int \ \Theta_{DMSO}(r) \  e^{-\beta W_{DMSO}(r)}  \  {\rm d}{\bf r} \right]
 \ \ \left[
\int \ (1-\Theta_{ACN}(r)) \   e^{-\beta W_{ACN}(r)} \  {\rm d}{\bf r}
\right]^2
   }    
\end{equation}
where $\Theta_{i}(r)$ is unity for those $r$ values which correspond 
to configurations of solvent $i$ exhibiting associated ion pairs;
 for other values $\Theta_{i}(r)$ is zero.
Equation (\ref{rough}) can be approximated by
\begin{equation}
{K_d(DMSO) \over K_d(ACN)} \simeq   \frac {  
\int_0 ^{r_{max}} \ e^{-\beta W_{ACN}(r)}  \ r^2 \ {\rm d} r  }{
\int_0 ^{r_{max}} \ e^{-\beta W_{DMSO}(r)}  \ r^2 \ {\rm d} r  }\simeq 
e^ {-\beta \Delta W(r_{max})}  \simeq \ 10^5
\end{equation}
where we have dropped terms of order one  and have assumed that
the boundary between associated and dissociated
ionic states lies at the first maximum beyond the CIP well.
Our simulations  not only predict the correct
experimental trend but also the result is in rough accord with
the large experimentally observed solvent effect on the acid
 ionization\cite{dis}, where 
$\left[K_d(DMSO) / K_d(ACN)\right]_{exp} =  6.9 \ 10^6$.
Finally, we note that the individual dissociation constants for both
solvents  differ by several orders of magnitude from the corresponding 
experimental values. This is expected in view of  
the simplifications introduced in the design of the potentials. In fact, 
differences in a few tenths of a kcal in the potentials lead to  
considerable variations in calculated dissociation constants.
Since our emphasis is on the study of relative, quantitative
changes in the ionization process as a function of solvent type and
environment, we have made no attempt to tune our simple potential model
to describe all quantitative aspects of the equilibrium structure.

\subsection{Bulk solvation structure}

The profiles of the
potential of mean force presented in  Fig.~1 show that there are 
no clearly identifiable solvent separated ion pairs for ACN, as
signaled by a secondary minimum in the mean potential. There is 
only a shallow minimum at $r_{s}=5.0$ \AA \ giving a barrier between the 
SSIP and CIP states of 
$\Delta W_{ACN}(bulk)=W_{ACN}(r_{max})-W_{ACN}(r_{s}) \simeq 1.0$ kcal/mol
which is within the statistical
uncertainty of our calculations.
 In contrast, the mean potential for DMSO shows a fairly deep 
secondary minimum at $r_{s}=7.5$ \AA \ with a barrier separating 
the CIP and SSIP states of 
$\Delta W_{DMSO}(bulk)=W_{DMSO}(r_{max})-W_{DMSO}(r_{s}) =5.4$ kcal/mol. 
In the DMSO  calculation, the curve  was computed from a constrained 
MD simulation as described above. The 
filled circles were determined using the formula
\begin{equation}
P(r) = P_u e^{-\beta W(r)}\;,
\end{equation}
where $P(r)$ is the probability density of $r$ and $P_u$ is the 
uniform distribution.
The probability density $P(r)$ was estimated 
from a histogram of the separation
between the ions obtained from a long unconstrained MD trajectory
initiated near the mean potential secondary minimum at $r_s=7.5$ \AA.
Note that most of the probability density is
confined between $6-10\;$ \AA \  with no escape to either contact ion pairs
or free ions observed in the course of the $4$ ns MD simulation.

An idea of the average solvation structure in DMSO can be obtained 
from Fig.~2, where we
show a contour plot  of the solvent charge density, 
$n_q(\rho,z)$,  for an interionic separation of $r=7.5$
\AA, which corresponds to SSIP separation
configuration. The charge density is defined as
\begin{equation}
\label{rrhhoo}
 \rho_{B} \  n_q(\rho,z) 
=\langle \sum_i \sum_\alpha \frac{q_\alpha}{2\pi \rho} 
\delta(\rho_{i}^{\alpha}-\rho)  
\delta(z_{i}^{\alpha}-z)  
\rangle_{r}\;,
\end{equation}
where $ \rho_{B}$ represents the solvent bulk density and  
the angular brackets $\langle \cdots \rangle_{r}$ denote 
an ensemble (or time) average over configurations with the ion
pair separation at a given value of   $r$. 
In (\ref{rrhhoo}),
we have selected a frame of reference 
centered on the ion pair, which is taken to lie along the $z$ direction in 
a cylindrical coordinate system $(\rho,z,\phi)$;
$(\rho_{i}^{\alpha},z_{i}^{\alpha})$ and
$q_\alpha$ represent the cylindrical coordinates and
charge of site $\alpha$ in the $i$-th solvent molecule, respectively.
The Cl$^-$ and H$^+$ ions are
denoted  by semicircular shaded regions in the bottom of
the figure while the $+$ and $-$ signs indicate regions of high
charge density of the corresponding sign. The picture provides information
about both the locations of high solvent density as well as the average
orientation of solvent molecules in the vicinity of the ion pair.
 Figure ~ 2  clearly shows that the solvation structure for DMSO
is energetically dominated by the electrical coupling between
the proton, a bare charge of small size, and the negatively charged 
oxygen sites of the solvent. 
This is indicated by the prominent ``$-++$"
charge density maxima to the right and above the proton; the 
molecular dipoles are oriented as might be expected, with the negative
ends of the molecules lying near to the H$^+$ side of the ion pair complex
and the positive ends of the other solvent molecules oriented
towards the Cl$^-$ side. However, the positive charge density near 
the Cl$^-$ ion is more diffuse indicating weaker solvation and less 
charge localization. 
The analysis of the solvation structure in ACN for a
similar ion pair separation shows essentially the same 
qualitative features as previously
described for DMSO, so a comparative analysis of the charge density
profiles is insufficient to understand the differences in 
ionic stabilization that these solvents present.   
Perhaps this is to be expected since the solvent contribution
to the potential of mean force is the result of a subtle interplay
between packing effects governed by short-range, repulsive forces 
related to the molecular shape,  and solute-solvent dipolar
forces which depend on overall distribution of  molecular
electronic density.    

The analysis of the solvent structure in the neighborhood of the
transition region between CIP and SSIP ion pair configurations 
presents interesting differences for the two solvents.
We examine the region   
where the steep attractive branch of the mean potential reaches its local 
maximum and levels off.
Consider the dipolar density,
\begin{equation}
 \rho_{B}  \ n_{\mu}(\rho,z) = \langle \sum_i 
 \frac{\hat{\mu}_i \cdot \hat{{\bf z}}}{2\pi \rho} \delta(\rho_{i}-\rho)  
\delta(z_{i}-z)  \rangle_{r}\;,
\end{equation}
in the $\rho z$-plane. Here $\mu_i$ is the dipole
moment of molecule $i$ and, as above, the average is taken with the distance
between the ions fixed at $r$.
Figures~3(a) and (b) show $n_{\mu}(\rho,z)$ for $r=2.64$ for ACN and
DMSO, respectively. For ACN one observes two large peaks with opposite
sign in the vicinity of H$^+$. This indicates that the ACN molecules have
a considerable degree of orientational variability so that when the
molecular dipole moment is projected onto the $z$ axis of the ion pair the
projection can take either sign. 
In contrast, the dipolar density for DMSO
in Fig.~3(b) shows no such effect implying that the DMSO molecules are
much more rigidly ordered in the first solvation shell as indicated in the
schematic representation of the structure shown in the bottom parts of the
figures. The results suggest that in ACN the local solvent density 
fluctuations are
sufficiently large to promote dipolar orientations that, in average,
lead to a less effective dielectric reactive field, lowering 
the energetic cost to pass from SSIP to CIP states. However, in DMSO 
a much more structured three-dimensional solvent network exists 
that hinders molecular rotations and leads  to a stronger
coulombic coupling between the proton and the negatively charged oxygen 
site and a larger free energy barrier.

\section{Acid Dissociation in clusters}
Compared to bulk environments, 
dissociation processes in small clusters present some distinctive features that may lead to 
significant variations in the overall dissociation energy of the ion pair.
Perhaps the simplest question to be considered is how large should a cluster  
be to exhibit bulk-like  behavior? The answer will depend on the particular
stage of the dissociation process one is interested in describing. 
In this study we focus our attention on the small-intermediate interionic 
distance regime; $i.e$, those characteristic of CIP and SSIP configurations 
seen in bulk phases. For interionic
distances which are comparable to typical linear dimensions of the cluster,  
surface forces introduce new effects in the dissociation 
mechanisms.\cite{laria}

In Fig.~4  we present the potential of mean force for HCl
dissociation in  clusters of ACN and DMSO. 
We first observe that the results for the mean potential for an $n=14$
ACN cluster, for interionic distances that are not too large, 
lie close to those for bulk solvents, indicating that even for this 
relatively small aggregate the solvent reactive
field is comparable to that of bulk environments and that surface 
effects play only a minor role in
determining the functional form of $W(r)$. 
Very small clusters exhibit
quite different behavior as is evident from the results for $W(r)$ for an
$n=4$ ACN cluster shown in the figure. The typical bulk phase plateau regime 
at moderate interionic distances has disappeared since 
for such small aggregates there is insufficient space  for independent
solvation of the individual ions.  
Figure~4  also shows results for the mean potential for 
a  DMSO cluster with $n=30$ at temperature $T=200\; K$. Our results confirm
that solvent separated ion pairs still persist in  
clusters of this size although the free energy barrier between the CIP and SSIP states
is smaller, $\Delta W_{DMSO}(n=30)=1.8$ kcal/mol.

Several interesting features of the dissociation mechanisms in ACN which 
are common to bulk 
and clusters environments  can be seen from typical snapshots of
cluster configurations. In Fig.~5  we present two often-encountered
cluster configurations for ACN  clusters with $n=30$ where the
ion pair separation is fixed at the transition  
value corresponding to the sharply rising part of the mean potential
discussed above.
The ACN molecules
lying closest to the H$^+$ ion are rendered in dark colors for contrast
while the remaining solvent molecules are lightly shaded. One sees that
while on average four ACN molecules strongly solvate the H$^+$ ion, their
orientations may be such that the projection of their dipole moments on
the $z$ axis can be positive or negative. In contrast,  DMSO clusters 
do not exhibit this dual dipolar orientation.

Conformational equilibria between clusters of distinctive
geometry occur as we approach the limiting regime where clusters split into 
two or more independent aggregates.\cite{laria}  
The interionic distance at which these phenomena are observed 
dependens on the cluster size.
For the very small $n=4$ ACN cluster case shown in Fig.~6 this occurs for 
an ion pair separation  $r=5.29$ \AA. At this ion pair separation one
sees the two types of solvent configurations;
namely, (a) three molecules strongly solvating the H$^+$ ion with the fourth
solvent molecule lying farther away and solvating the Cl$^-$ ion. The
other prominent configuration (b) occurs when all four solvent molecules
strongly solvate the H$^+$ ion. This is a transitional distance: for
smaller separations one always sees the solvation structure given by (a)
while for larger separations one always sees that corresponding to (b). 
If one considers larger clusters analogous phenomena exist, however here
we have to consider larger ion pair separations. As an example, Fig.~7 shows 
results for a $n=14$ ACN cluster with
the ion pair separation fixed at the very large value $r=27$ \AA. The
equilibrium structure is determined by configurations (a) where varying
numbers of solvent molecules solvate the H$^+$ and Cl$^-$ ions but these
two solvated ions form subclusters which only weakly interact.
However, there also exist equilibrium configurations (b) where long ``strings"
of solvent molecules form a link or ``rubber band" connecting the two ions
leading to a significant attractive force. This is a distinct cluster
effect which has no counterpart in the bulk. Such structures were observed
earlier for ions of like charge in water.\cite{laria} However, we remark that in the regime of
ion pair separations investigated there is no hysteresis in the mean
potential indicating that the equilibrium configuration space is being
sampled fully even for the ``rubber band" configurations shown in
Fig.~7.

Two examples of cluster structure with the ion pair at $r=7.5$ \AA, the
SSIP configuration for DMSO, are shown in Fig.~8, one for ACN (top) and
the other for DMSO (bottom). The noteworthy features of both panels are
the obvious strong orientational order near H$^+$ for both solvent species.
However note that  although 
 there is little discernible difference between the
two solvent cases, SSIPs exist for clusters of DMSO but do not for ACN.
These two pictures further illustrate the fact that the cluster shape
fluctuations are strong and these fluctuations must be taken into account
in any model of the cluster solvation dynamics. Furthermore, one sees that
surface forces do manifest themselves in the orientational order of the
solvent molecules on the cluster surface which tend to lie parallel to
the surface. However, as observed earlier, even for fairly small clusters
($n=14$) there are only rather small quantitative effects on the mean potential
and such surface force effects become important only at longer ion pair
separations or for smaller cluster sizes.

\section{Conclusions}
The calculations of the ionization of HCl in ACN and DMSO solvents 
 indicated some of the solvation features that make the
ionization process rather different in these solvents in spite of their
similar dipole moments. The differences could be ascribed to the details
of the charge distributions in the molecules and their effects on the
quantitative aspects of the solvation structure. The gross features of
the solvation structure were shown to be similar for both solvents.
However, in spite of the apparent similarities in the local solvation
structure, DMSO supports solvent separated ion pairs while the tendency to
form such pairs is considerably reduced or absent in ACN.

The cluster environment has an important influence on the ionization 
process, especially for clusters with fewer than $n=14$ molecules. For
ion pair separations as large as that for the SSIP state, the cluster potential of
mean force closely corresponds to that of the bulk, even for clusters with as 
few as $n=14$ solvent molecules. For much smaller clusters, as might be
expected, there are significant changes in the mean potential reflecting
the finite size of the cluster environment. Of course, as the ion pair
is stretched the finite size of the cluster gives rise to much more dramatic
effects on the mean potential as reflected in the ``rubber band"-like 
solvent configurations that lead to attractive forces between the ions
that cannot exist in the bulk. Since these forces correspond to an
equilibrium cluster and involve averages over all cluster configurations
with  the ion pair separation fixed, they may have little relevance for 
ion pair dissociation processes in clusters since the time scale to
establish such equilibrium structures may be long compared to typical
times for dissociation. It would be interesting to investigate the
dynamics of dissociation in clusters in this context.

In order to estimate the importance of the explicit incorporation of the
quantum nature of the proton in our calculations, we performed a series
of simulations in an $n=14$ ACN cluster where the proton was treated 
as a classical point charge. The results for the computed potential
of mean force  are included in Fig.~4; they 
 clearly show the necessity  
to treat the proton quantum mechanically in order to
accurately represent the activation free energy for ionization in these
solvents. 
Classical mechanics underestimates the barrier for passage from 
the CIP to SSIP or free ion states.  In this respect, the present behavior
is opposite to that found in proton transfer reactions where quantum
tunneling reduces  the height of the effective
free energy barrier.\cite{path}
Quantum dispersion tends to delocalize the proton charge  leading
to a less effective ionic solvation, as well as smaller solvation energies
for dissociated ions  in comparison
to results from  classical models where the
proton is taken to be a localized point charge.

The simulations presented in this paper show that acid ionization in
molecular clusters can mimic that in the bulk even for quite small
clusters. However, for clusters below a certain size, about $n=14$ in
this study, distinctive finite-size effects influence the ionization.
The study points to some of the new chemistry that can occur as a result
of environmental changes on the microscopic and mesoscopic levels.

\begin{center}
{\bf Acknowledgments}
\end{center}
D.L. is a recipient of a NSERC Foreign Researcher Award.
The research of R.K. was supported in part by a 
grant from the Natural Sciences and Engineering Research Council of Canada.
This research was also partially supported 
by the EEC Network Contract ERBCHRXCT930351 "Molecular Dynamics and
Monte Carlo simulations of quantum and classical systems". 
We finally thank M. Ferrario for stimulating discussions.

\newpage
\begin{figure}[htbp]
\epsfxsize = \textwidth
\epsfbox{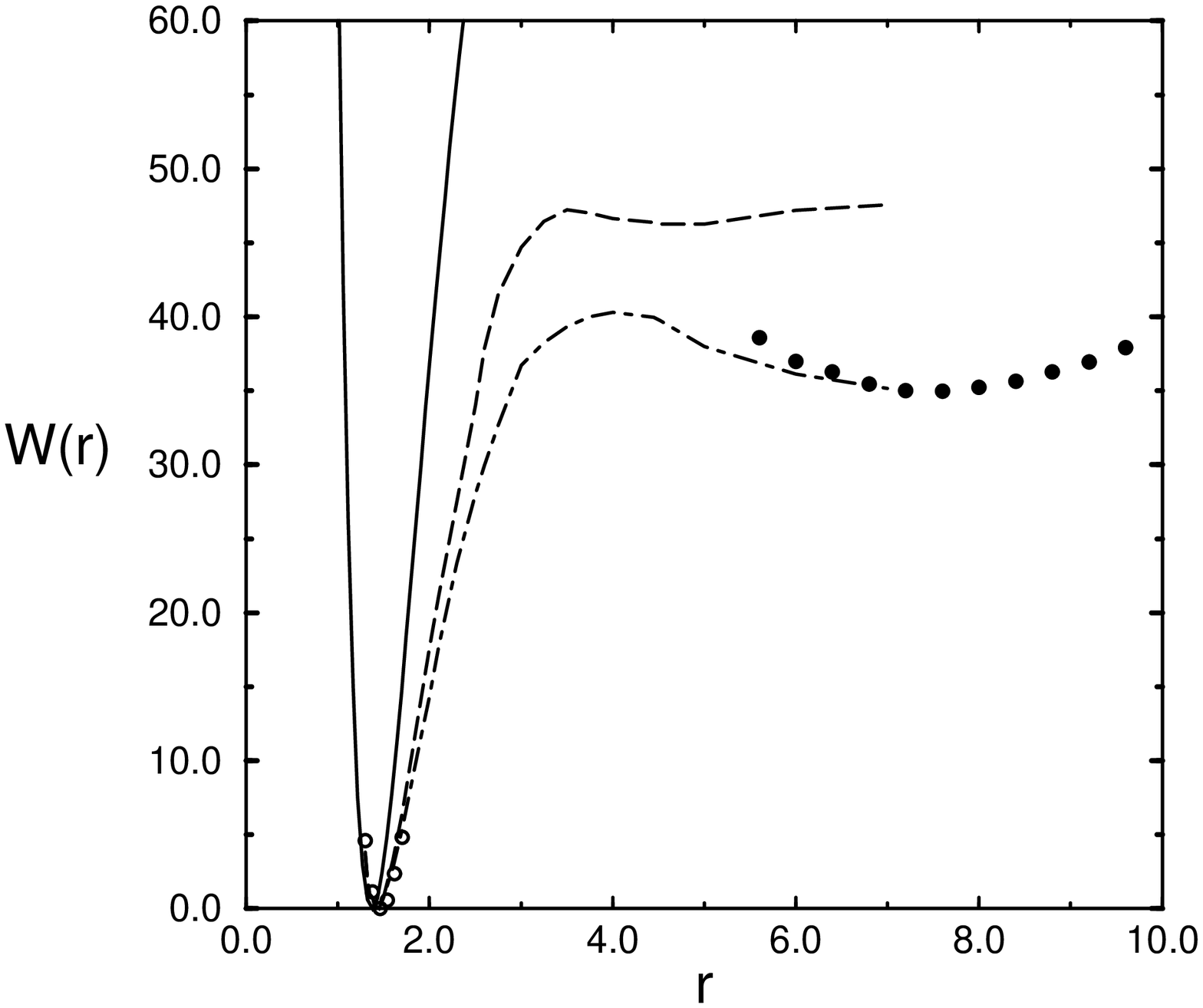}
\caption{Potential of mean force in units of kcal/mol for bulk phase 
systems: solid line, bare potential; dashed lines, ACN; dashed-dotted 
lines, DMSO; open circles correspond
to unconstrained trajectories results for DMSO and ACN at small
distances; filled circles correspond to unconstrained
trajectories results for DMSO at large distances. The distance
$r$ is in units of \AA.}
\label{Fig. 1}
\end{figure}

\newpage
\begin{figure}[htbp]
\epsfxsize = \textwidth
\epsfbox{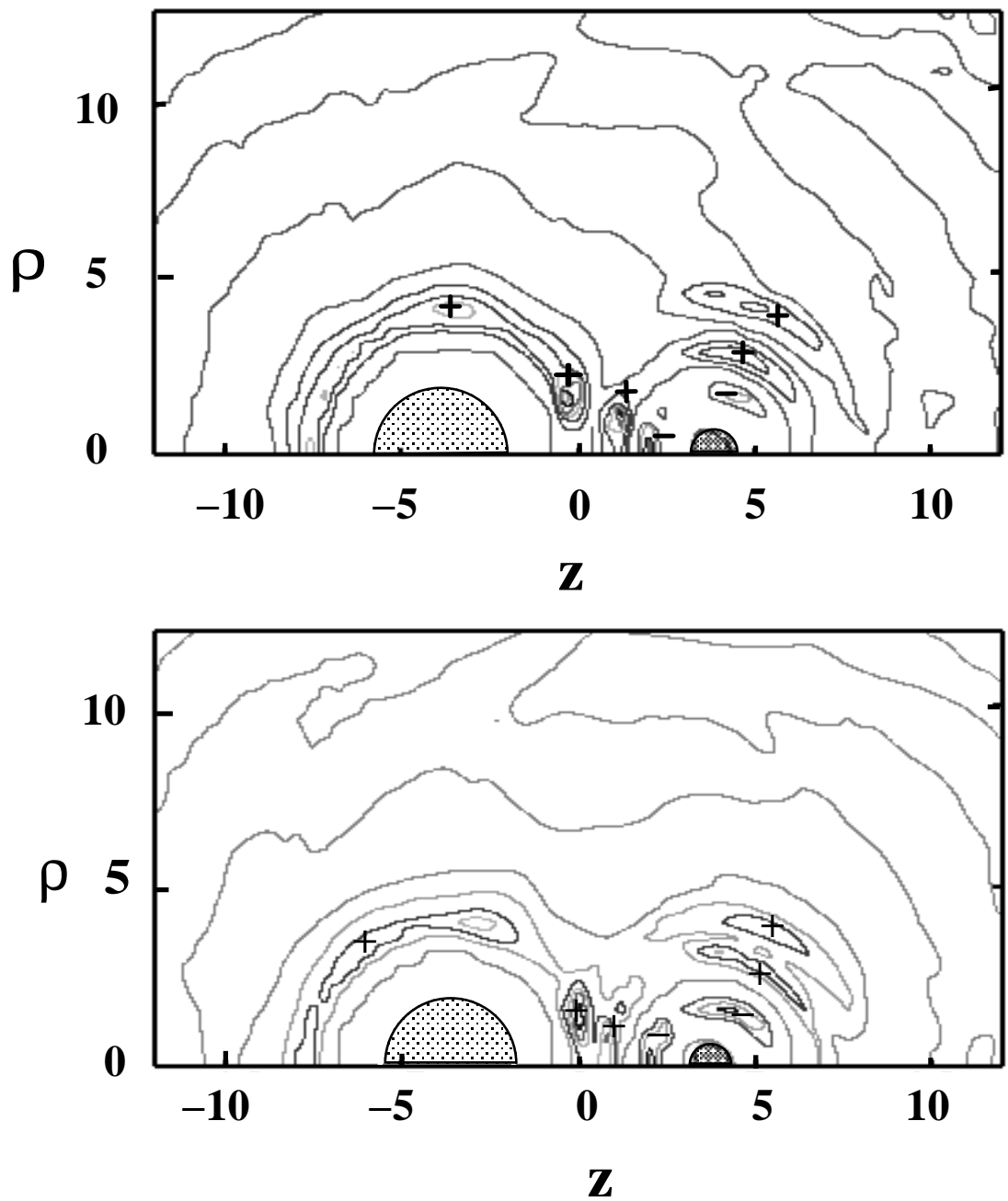}
\caption{Contour plots of the charge density $n_{q}$ in the 
$\rho z$-plane with the ion pair in the SSIP configuration at $r=7.5$ \AA. 
Top and bottom  panels correspond to DMSO and ACN, respectively. }
\label{Fig. 2}
\end{figure}

\newpage
\begin{figure}[htbp]
\epsfxsize = \textwidth
\epsfbox{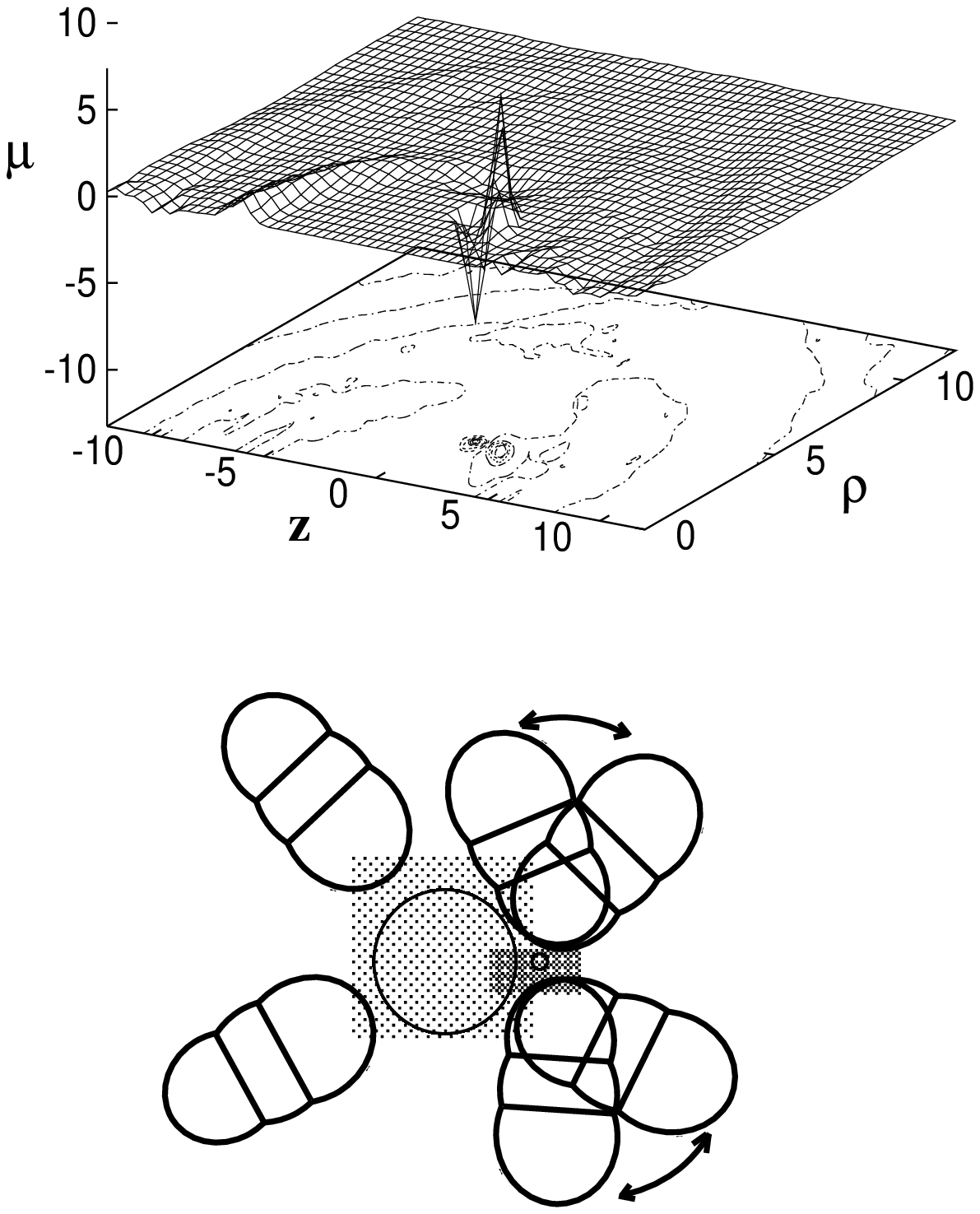}
\end{figure}

\newpage
\begin{figure}[htbp]
\epsfxsize = \textwidth
\epsfbox{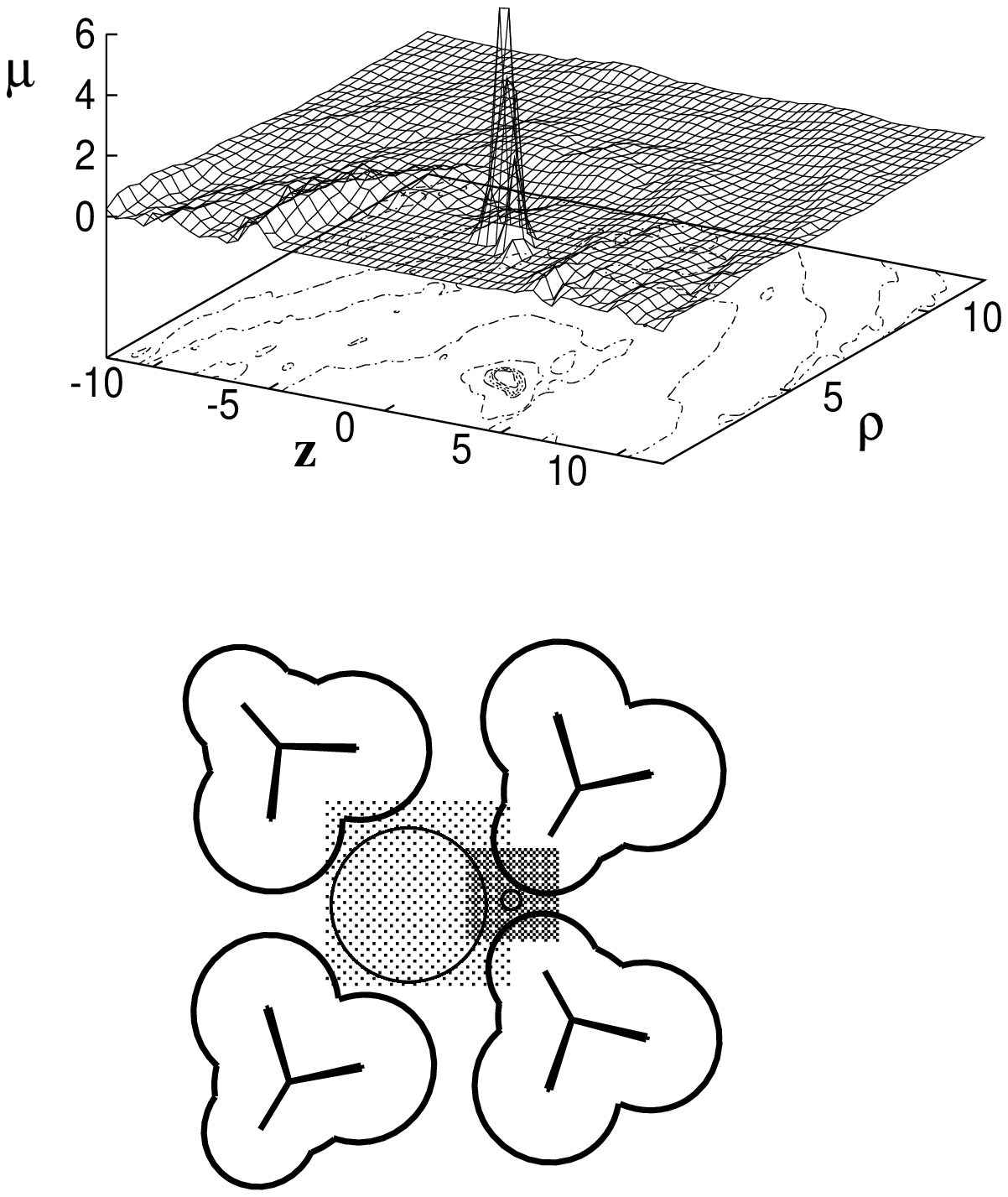}
\caption{Dipolar density $n_{\mu}$ (written as $\mu$ on the ordinate) 
with the ion pair at $r=2.64$ \AA: (a) ACN and (b)
DMSO. Schematic representations of the local solvent structure are 
shown in the botton parts of the figures.}
\label{Fig. 3}
\end{figure}

\newpage
\begin{figure}[htbp]
\epsfxsize = \textwidth
\epsfbox{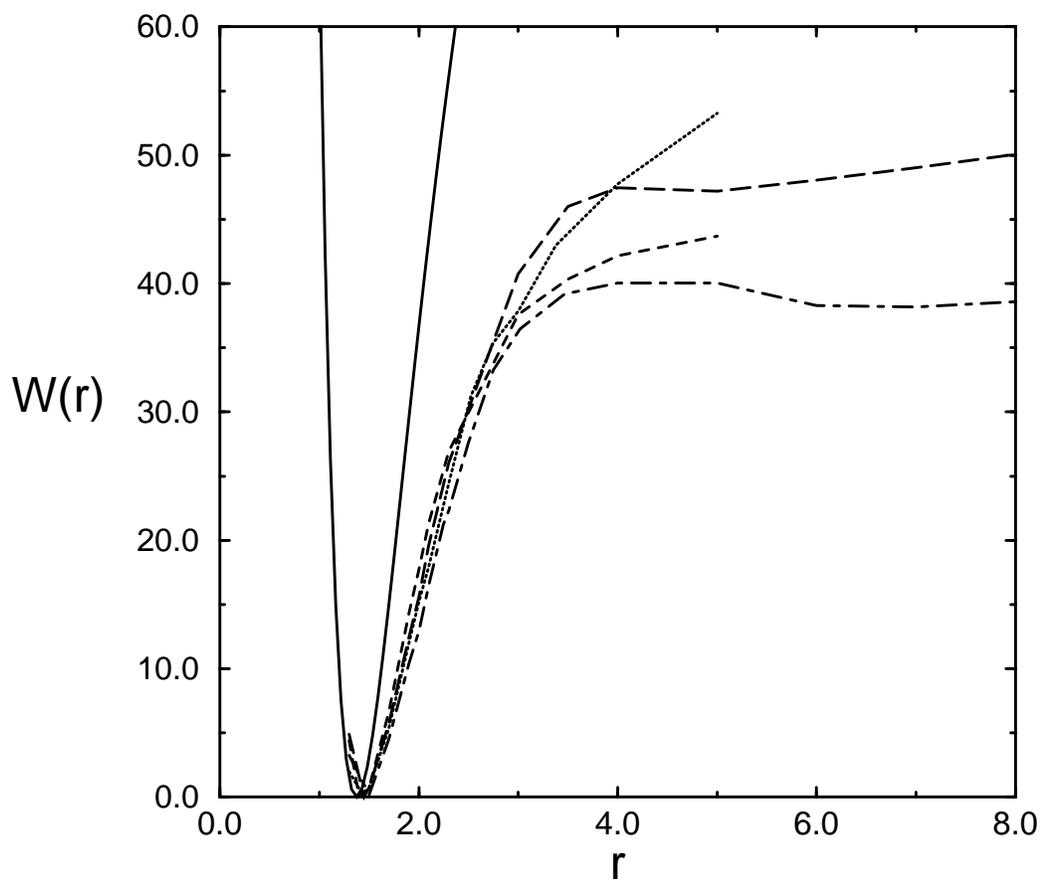}
\caption{Potential of mean force for clusters: solid line, bare potential; 
dashed line, ACN cluster $n=14$; dotted line, ACN cluster $n=4$;
dotted-dashed line, DMSO cluster $n=30$; short-dashed lines, ACN cluster
$n=14$ treating the proton classically. Units are the same as in Fig. 1.} 
\label{Fig. 4}
\end{figure}

\newpage
\begin{figure}[htbp]
\epsfxsize = \textwidth
\epsfbox{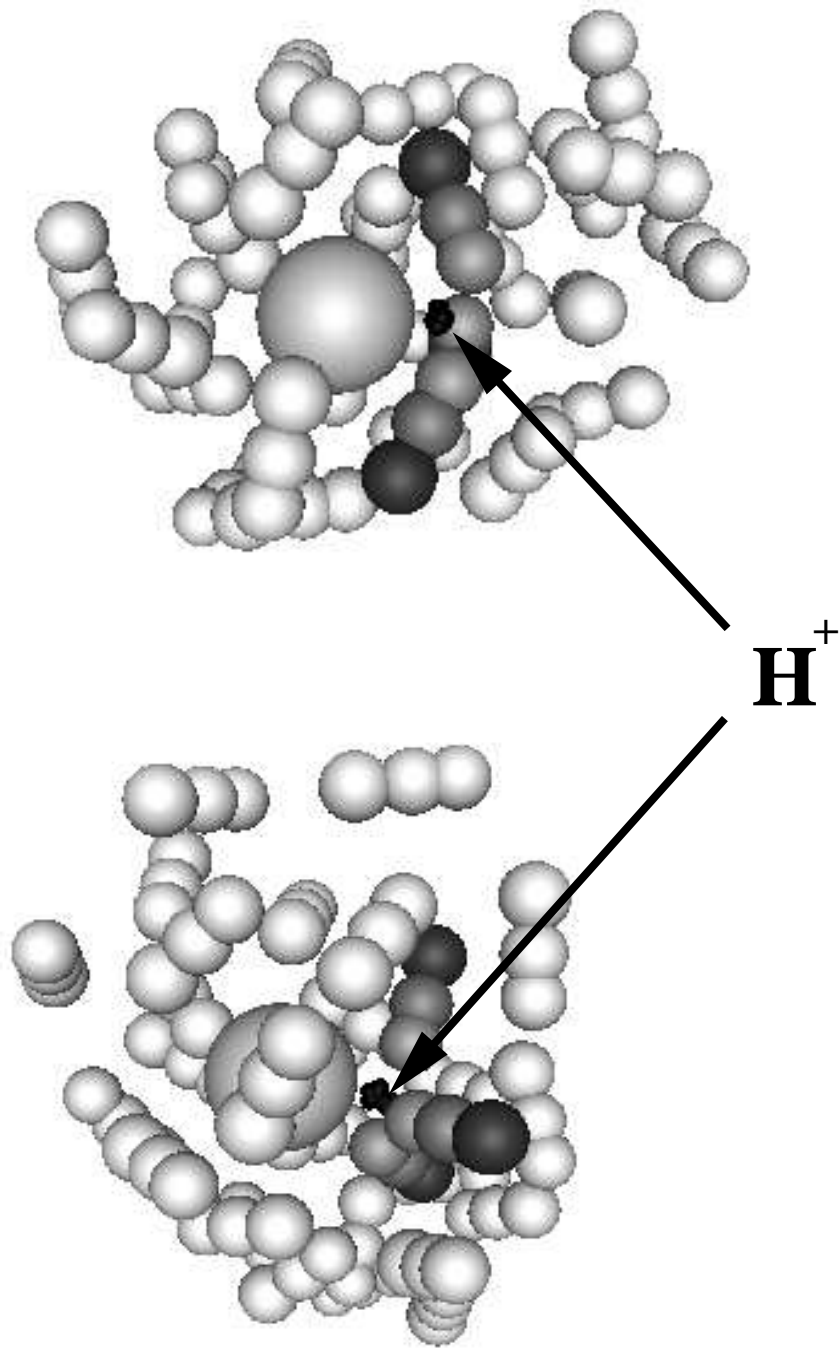}
\caption{Two configurations of an $n=30$ ACN cluster with $r=2.64$ \AA. The 
molecules strongly solvating H$^+$ are heavily shaded and the two
configuration illustrate the different orientations of the solvent
molecules around H$^+$.}
\label{Fig. 5}
\end{figure}

\newpage
\begin{figure}[htbp]
\epsfxsize = \textwidth
\epsfbox{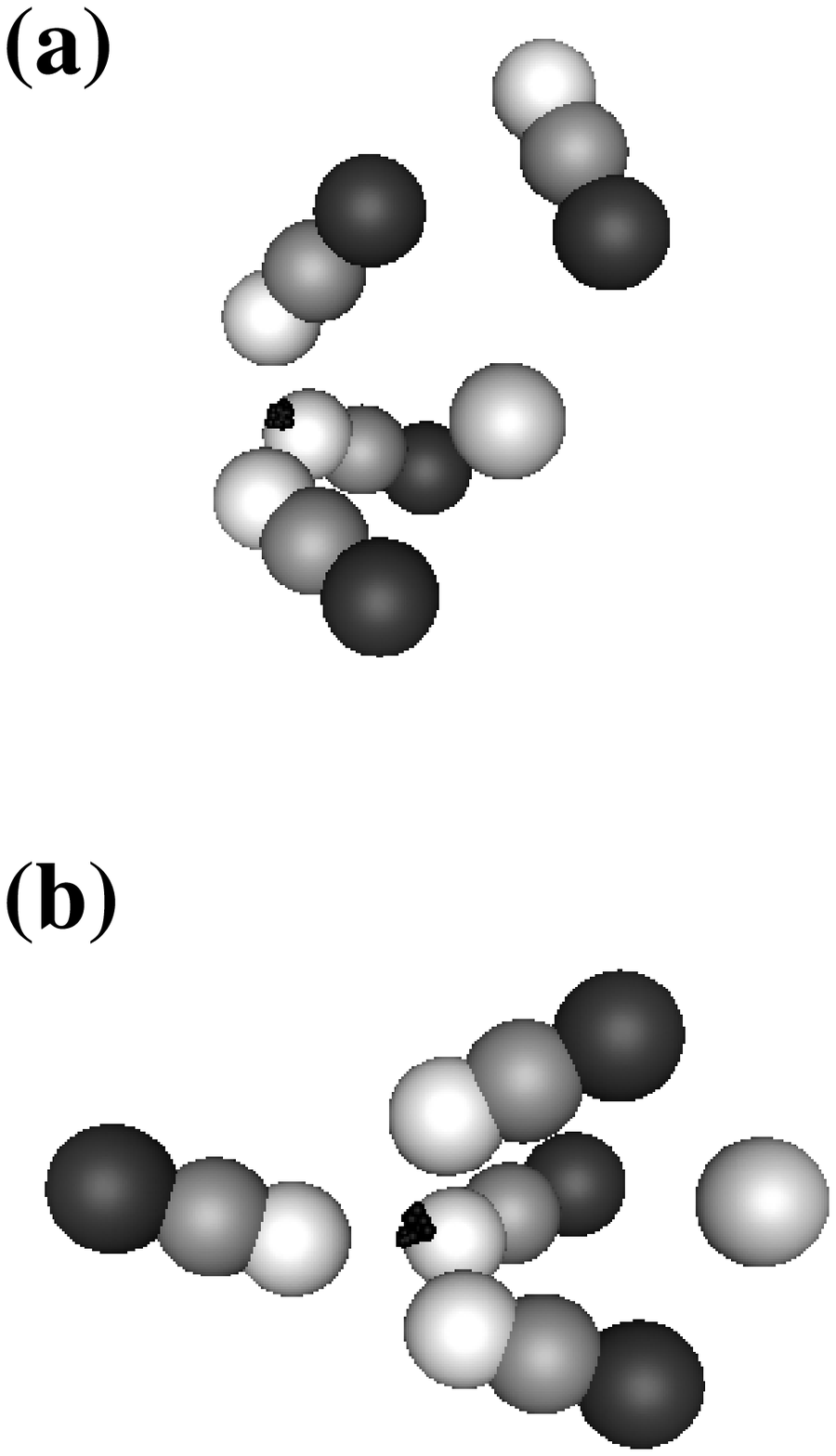}
\caption{ACN cluster with $n=4$ with the ion pair fixed at $r=5.29$ \AA.
Parts (a) and (b) show two different metastable configurations.}
\label{Fig. 6}
\end{figure}

\newpage
\begin{figure}[htbp]
\epsfxsize = \textwidth
\epsfbox{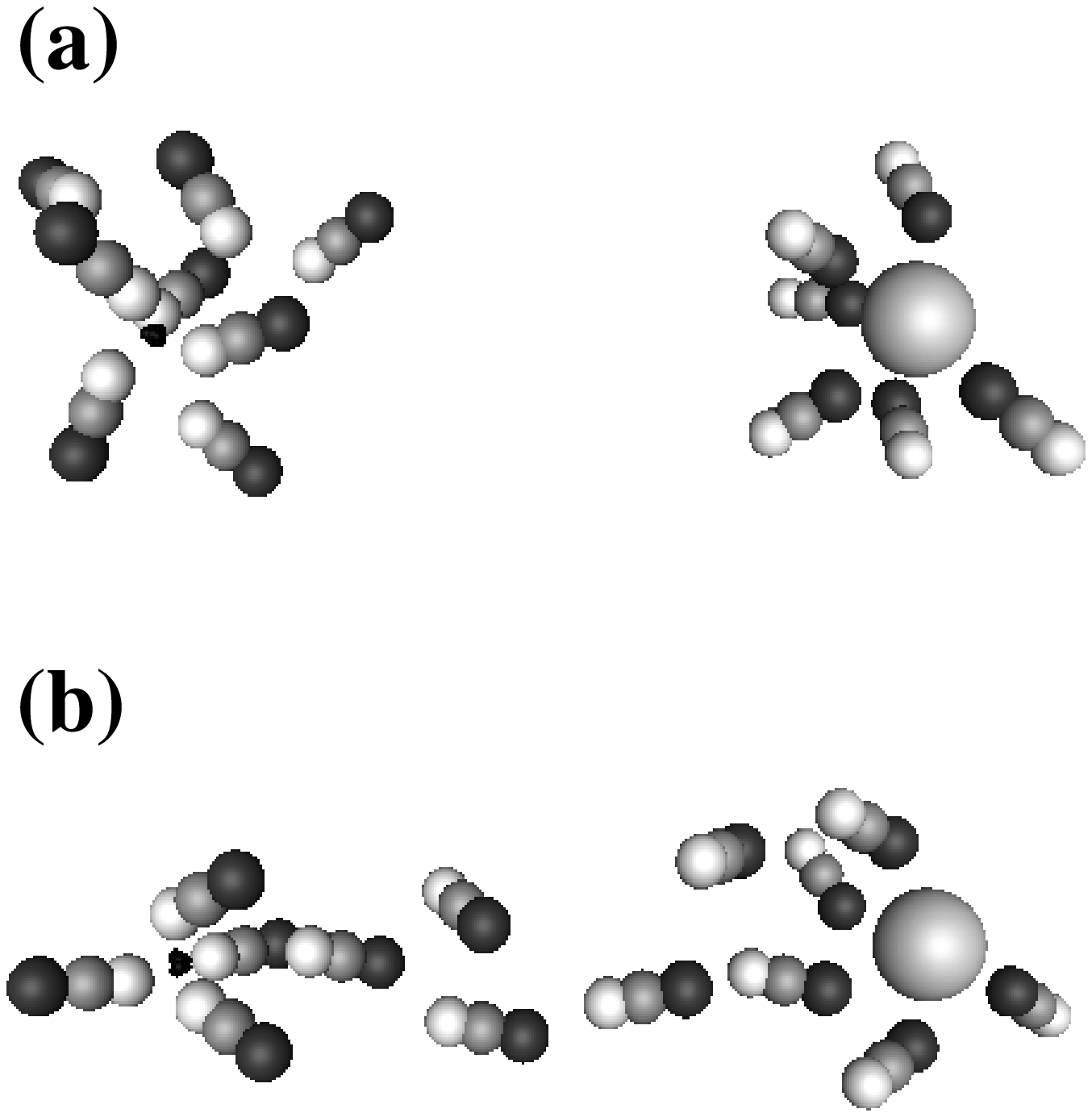}
\caption{ACN cluster with $n=14$ and the ion pair distance fixed at 
$r=27$ \AA. Parts (a) and (b) are two prominent metastable configurations.}
\label{Fig. 7}
\end{figure}

\newpage
\begin{figure}[htbp]
\epsfxsize = \textwidth
\epsfbox{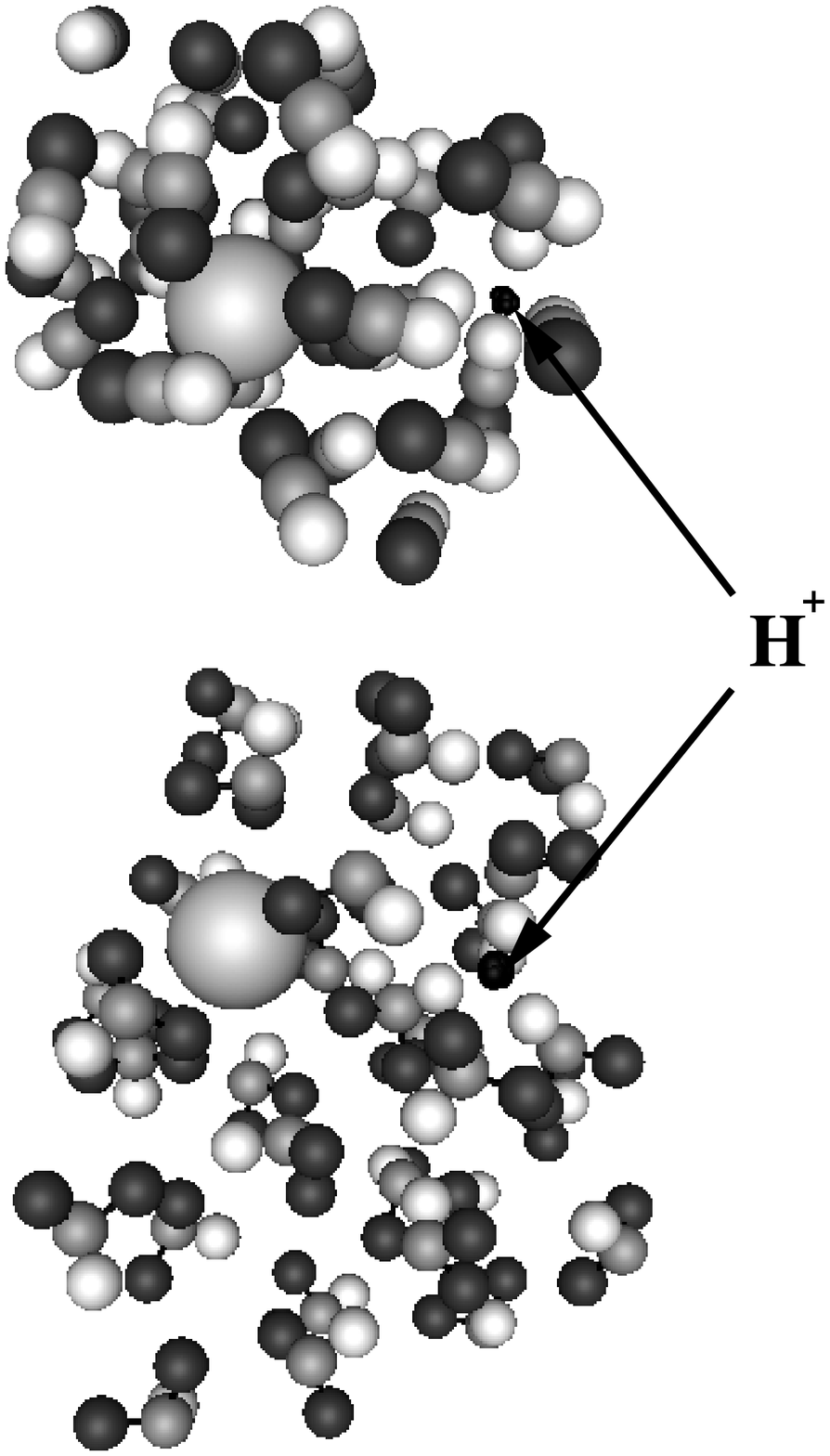}
\caption{Two clusters with $n=30$ showing the solvent structure 
with the ion pair distance fixed at $r=7.5$ \AA. Top, ACN; bottom, DMSO.}
\label{Fig. 8}
\end{figure}


\begin{thebibliography}{99}

\bibitem{dis} C. Cooke, C. McCallum, A.D. Pethybridge and J.E. Prue,
 Electrochim. Acta, {\bf 20}, 591 (1975). I.M. Kolthodd, S. Bruchenstein and 
M.K. Chantooni, Jr. J. Am. Chem. Soc. {\bf 83}, 3927 (1961).
\bibitem{grot} For a discussion of this mechanism see, for example, S.
Glasstone, {\em Physical Chemistry}, (D. Van Nostrand Company, New York,
1959). 

\bibitem{hynes} For a recent theoretical study of such effects for HCl
ionization in water see, K. Ando and J.T. Hynes, J. Am. Chem. Soc., to 
be published.


\bibitem{ssip} S. Winstein, E. Clippenger, A.H. Fainber and G.C.
Robinson, J. Am. Chem. Soc. {\bf 76}, 259 (1954).

\bibitem{ipsim} Simulations of SSIP configurations for classical ions in
various model solvents have been carried out. See, for instance, O.A.
Karim and J.A. McCammon, J. Am. Chem. Soc. {\bf 108}, 1762 (1986); Chem.
Phys. Lett. {\bf 132}, 219 (1986); G. Ciccotti, M. Ferrario, J.T. Hynes
and R. Kapral, Chem. Phys. {\bf 129}, 241 (1989); J. Chem. Phys. {\bf
93}, 7137 (1990).


\bibitem{ct}
O. Cheshnovsky and S. Leutwyler, Chem. Phys. Lett.
{\bf 121}, 1 (1985);
R. Knochenmuss, O. Cheshnovsky and S. Leutwyler, $ibid$,
{\bf 144}, 317 (1988).

\bibitem{ct1}
J. A. Syage and J. Steadman,  J. Chem. Phys.
{\bf 95}, 2497 (1991).

\bibitem{cast}
B. D. Kay, V. Hermann and A. W. Castleman, Jr., Chem. Phys. Lett.
{\bf 80}, 469 (1981).


\bibitem{shake} J.P. Ryckaert, G. Ciccotti and H. J. C.
Berendsen, J. Comput. Phys., {\bf 23}, 327 (1977).

\bibitem{edwards} D.M.F. Edwards, P.A. Madden and I.R. McDonald, Mol.
Phys. {\bf 51}, 1141 (1984).

\bibitem{feder} W. Feder, H. Dreizler, H.D. Rudolph and V. Typke, 
Z. Naturforsch. {\bf 24A}, 266 (1969).

\bibitem{rao} B.G. Rao and U.C. Singh, J. Am. Chem. Soc. {\bf 112}, 3802 (1990).

\bibitem{dft1} S.H. Vosko, L. Wilk and M. Nusair. Can. J. Phys. {\bf 58}, 1200
 (1980). A.D. Becke. Phys. Rev. A {\bf 38}, 3098 (1988). J.P. Perdew. Phys. Rev.
B {\bf 33}, 8800 (1986).

\bibitem{dft2} E. Clementi, S.J. Chakravorty, G.Corongiu, J.R. Flores and
V. Sonnad. $MOTECC91$, Chapt. 2. edited by E. Clementi. (ESCOM, Leiden, 1991).

\bibitem{rossky} B.M. Pettitt and P. Rossky, J. Chem. Phys. {\bf 84}, 5836 (1986).

\bibitem{feyn} R. P. Feynman, {\it Statistical Mechanics} (Addison
Wesley, Reading, 1972).

\bibitem{chand} D. Chandler and P. Wolynes, J. Chem. Phys. {\bf 74}, 
4078 (1981).

\bibitem{nose} S. Nos\'{e}, Mol. Phys. {\bf 57}, 187 (1986)

\bibitem{verlet} L. Verlet, Phys. Rev. {\bf 159}, 98 (1967).

\bibitem{path} The calculation is similar to that for proton transfer: 
D.Laria, G.Ciccotti, M.Ferrario and R.Kapral, Chem. Phys.
{\bf 180}, 181 (1994).

\bibitem{gillan} M. Gillan, J. Phys. C {\bf 20}, 3621 (1987).
\bibitem{voth} G. A. Voth, D. Chandler and W. H. Miller, J. Chem.
Phys. {\bf 91}, 7749 (1989).

\bibitem{crcd} E. A. Carter, G. Ciccotti, J. T. Hynes and R. Kapral,
Chem. Phys.  Lett. {\bf 156}, 472 (1989).

\bibitem{laria} D. Laria and R. Fern\'andez-Prini, J. Chem. Phys., 
{\bf 102},    (1995).

\end{thebibliography}
\end{document}